%% file: main.tex
\RequirePackage{fix-cm}
\documentclass[smallextended]{svjour3}       
\smartqed  

\usepackage{wrapfig}
\usepackage{tabularx}
\usepackage{makecell}
\usepackage{multirow}
\usepackage{subfigure}
\usepackage{algorithm}
\usepackage{algorithmicx}
\usepackage{amsmath,amsfonts}
\usepackage{amssymb}
\usepackage{array}
\usepackage{balance}
 \usepackage{booktabs}
\usepackage[skip=1pt,labelfont=bf]{caption}
 \usepackage{calligra}
\usepackage{color}
\usepackage{colortbl}
\usepackage{courier}
\usepackage{csvsimple}
\usepackage{enumitem}
\usepackage{fancybox}
\usepackage{fontenc}
\usepackage{graphicx}
\usepackage{listings}
\usepackage{longtable}
\usepackage{lscape}
\usepackage{makecell}
\usepackage{moreverb}
\usepackage{multicol}
\usepackage{multirow}
\usepackage{pifont}
\usepackage{rotating}
\usepackage{setspace}
\usepackage{subfigure}
\usepackage[most]{tcolorbox}
\usepackage{threeparttable}
\usepackage{tikz}
\usepackage{soul}
\usepackage[normalem]{ulem}
\usepackage{url}
\usepackage{soul}
\usepackage{wasysym}
\usepackage{xspace}
\usepackage{hyperref}
\usepackage[misc]{ifsym}

\newcommand*{\affaddr}[1]{#1} 
\newcommand*{\affmark}[1][*]{\textsuperscript{#1}}

\algnewcommand\algorithmicforeach{\textbf{for each}}
\algdef{S}[FOR]{ForEach}[1]{\algorithmicforeach\ #1\ \algorithmicdo}

\newcolumntype{L}[1]{>{\raggedright\let\newline\\\arraybackslash\hspace{0pt}}m{#1}}
\newcolumntype{C}[1]{>{\centering\let\newline\\\arraybackslash\hspace{0pt}}m{#1}}
\newcolumntype{R}[1]{>{\raggedleft\let\newline\\\arraybackslash\hspace{0pt}}m{#1}}

\definecolor{codegreen}{rgb}{0,0.6,0}
\definecolor{codered}{rgb}{1,0,0}
\definecolor{codegray}{rgb}{0.5,0.5,0.5}
\definecolor{codepurple}{rgb}{0.58,0,0.82}
\definecolor{backcolour}{rgb}{0.95,0.95,0.92}
\definecolor{lightgray}{gray}{0.9}

\lstdefinestyle{mystyle}{
    commentstyle=\color{codegreen},
    keywordstyle=\color{magenta},
    numberstyle=\small\color{black},
    stringstyle=\color{codepurple},
    basicstyle=\scriptsize\ttfamily,
    breakatwhitespace=false,
    breaklines=true,
    captionpos=b,
    keepspaces=true,
    showspaces=false,
    showstringspaces=false,
    showtabs=false,
    tabsize=2
}

\lstdefinelanguage{diff}{
  morecomment=[f][\color{blue}]{@@},     
  morecomment=[f][\color{red}]-,         
  morecomment=[f][\color{codegreen}]+,       
  morecomment=[f][\color{red}]{---}, 
  morecomment=[f][\color{codegreen}]{+++},
}

\setlist{noitemsep} 

\definecolor{darkpastelred}{rgb}{0.76, 0.23, 0.13}
\definecolor{ao(english)}{rgb}{0.0, 0.5, 0.0}

\definecolor{darkpastelred}{rgb}{0.76, 0.23, 0.13}
\definecolor{ao(english)}{rgb}{0.0, 0.5, 0.0}

\definecolor{yellow}{RGB}{255,255,153}
\definecolor{grey}{RGB}{224,224,224}

\newboolean{showcomments}
\setboolean{showcomments}{true}
\ifthenelse{\boolean{showcomments}}
 { \newcommand{\mynote}[2]{
      \fbox{\bfseries\sffamily\scriptsize#1}
        {\small$\blacktriangleright$\textsf{\emph{#2}}$\blacktriangleleft$}}}
        { \newcommand{\mynote}[2]{}}

\definecolor{DarkOrange}{rgb}{0.8,0.3,0.0}
\definecolor{DarkCyan}{rgb}{0.0, 0.55, 0.55}
\definecolor{DarkCyel}{rgb}{1.0, 0.49, 0.0}
\definecolor{yellow-green}{rgb}{0.6, 0.8, 0.2}

\newcolumntype{?}{!{\vrule width 1pt}}

\newcommand{\tool}{\textsc{BugRMSys}\xspace}

\newcommand{\find}[1]{
\begin{tcolorbox}[leftrule=1mm,toprule=0mm,bottomrule=0mm,left=1pt,right=2pt,top=2pt,bottom=2pt]
\em #1
\end{tcolorbox}
}

\begin{document}

\title{App Review Driven Collaborative Bug Finding
}


\author{
    Xunzhu~Tang\protect\affmark[1] 
\and 
    Haoye~Tian\protect\affmark[1] 
\and
    Pingfan~Kong\protect\affmark[1] 
\and
    Kui~Liu\protect\affmark[2] 
\and
    Xin~Xia\protect\affmark[2]
\and
    Jacques~Klein\protect\affmark[1]
\and
    Tegawend\'e~F.~Bissyand\'e\protect\affmark[1]
}


\institute{
Xunzhu Tang \\ \email{xunzhu.tang@uni.lu}  \\\\
\Letter $\;\;$~Haoye~Tian\\ \email{haoye.tian@uni.lu}  \\\\
Pingfan Kong \\ \email{fandsec@gmail.com} \\\\
Kui Liu \\ \email{brucekuiliu@gmail.com} \\\\
Xin Xia \\ \email{xin.xia@acm.org} \\\\
Jacques~Klein \\ \email{jacques.klein@uni.lu} \\\\
Tegawend\'e~F.~Bissyand\'e \\ \email{tagewende.bissyande@uni.lu} \\\\
\affaddr{\affmark[1] SnT, University of Luxembourg, Luxembourg City, Luxembourg. \\
\affmark[2] Huawei, Hangzhou City, China.}\\
}

\date{Received: date / Accepted: date}

\maketitle

\input{0.Abstract.tex}
\input{1.Introduction}

\input{2.Motivating}

\input{3.Approach}
\input{4.Emperical_design}
\input{5.Emperiment_results}
\input{6.Discussion}
\input{2.Relatedwork}

\input{7.Conclusion}

\balance
\bibliographystyle{spmpsci}
\bibliography{bib/references.bib}
\end{document}

%% file: 0.Abstract.tex
\begin{abstract}
Software development teams generally welcome any effort to expose bugs in their code base.
In this work, we build on the hypothesis that mobile apps from the same category (e.g., two web browser apps) may be affected by similar bugs in their evolution process. It is therefore possible to transfer the experience of one historical app to quickly find bugs in its new counterparts. This has been referred to as collaborative bug finding in the literature. Our novelty is that we guide the bug finding process by considering that existing bugs have been hinted within app reviews. Concretely, we design the \tool approach to recommend bug reports for a target app by matching historical bug reports from apps in the same category with user app reviews of the target app. We experimentally show that this approach enables us to quickly expose and report dozens of bugs for targeted apps such as Brave (web browser app). 
\tool's implementation relies on DistilBERT to produce natural language text embeddings. Our pipeline considers similarities between bug reports and app reviews to identify relevant bugs. We then focus on the app review as well as potential reproduction steps in the historical bug report (from a same-category app) to reproduce the bugs.
Overall, after applying \tool{} to six popular apps, we were able to identify, reproduce and report 20 new bugs: among these, 9 reports have been already triaged, 6 were confirmed, and 4 have been fixed by official development teams, respectively.

\keywords{Bug finding \and App review \and Bug similarity \and Bug report}
\end{abstract}

%% file: 1.Introduction.tex
\section{Introduction}\label{sec:intro}

Modern apps must evolve quickly to adapt to a fierce competition in app markets where users have varied choices among feature-rich apps~\cite{mcilroy2016fresh}.
Unfortunately, the fast iteration in app updates often results in defects being found by users after releases~\cite{calcagno2015moving}.
Various research efforts based on static analysis~\cite{jiang2017detecting,lee2016hybridroid,talukder2019droidpatrol} and dynamic testing~\cite{hu2014efficiently,van2013dynamic,fan2018large,su2020my,liu2022guided} have therefore been carried out to detect bugs before releasing apps. Bug-free apps remain however a myth and even popular apps, which are intensively used by large user communities, often display simple but annoying defects~\cite{fan2018large,amalfitano2018does,sun2021understanding}. Through app reviews, users can provide feedback on buggy behaviour that sometimes go overlooked by app developers for various reasons: reviews can be redundant and uninformative (e.g. simple praise or dispraise repeating the star rating)~\cite{maalej2016automatic}. App reviews are also time-consuming to exploit and can mislead the identification of fault locations~\cite{stanik2019classifying}. In contrast, official bug reports filed in the issue tracker are the focus of developer communities since these reports tend to be more readily exploitable for bug resolution. 

It is noteworthy that if recurring bugs are not swiftly addressed by developers, they will lead to negative app reviews with significant impact on app score in app markets and other severe consequences such as app fails~\cite{li2010user}. 
The aforementioned situation calls for a more careful consideration of user reviews by developers. In particular, it would be appealing to translate app reviews into bug reports that can be used by developers as starting points in their fight against bugs.
However, there exists a significant gap between the language of user reviews and the language of developer bug reports. The former is generally formal and technically-written while the latter is informal and colloquially-written. In a recent work, Haering et al.~\cite{haering2021automatically} proposed a deep learning approach to match app reviews and bug reports with the ambition of easily tracking whether an issue reported in app review was already filed as an official bug report, which should increase bug fixing priority. While we subscribe to the claim that user feedback often lacks information that is relevant for developers (such as steps to reproduce or affected versions)~\cite{martens2019extracting,zimmermann2010makes}, their approach (1) does not address the key problem of review deluge, and (2) misses the opportunity to reveal new bugs to the developers. Indeed, on the one hand, for a popular app, there can be thousands of new reviews every day, most of which are noisy for developers since they do not offer insights into bugs. On the other hand, some app reviews may actually mention important and annoying bugs which can impact user experience for a large number of users without ever being reported formally in the issue tracker. 

In another research direction, Tan et al.~\cite{tan2020collaborative} have proposed Bugine~\cite{tan2020collaborative}, a collaborative bug recommendation system that aims at pairing similar issue reports across different apps. 
Thanks to Bugine, they have empirically shown that it is indeed possible to match similar issue reports across different apps. 
However, Bugine can only report issues across apps where the relevant UI design is of high visual similarity. Besides limited to only UI-related bugs, Bugine does not take target app's review into consideration, which enable it no ability to pick up useful bugs as input.

Following up on the hypothesis of the work by Tan et al., we performed a preliminary study (cf. Section~\ref{sec:motivating}) to check whether apps in the same category (e.g., two web browser apps or two calendar apps) tend to face similar development issues.
Eventually, we observed that apps in the same category share similar issues since these apps (1) are built by leveraging the similar development frameworks for similar functionalities (e.g., \textit{Unity} for gaming apps), (2) have similar UI design logic, and (3) use the same storage/notification/hosting services (e.g., FireBase)~\cite{long2014enabling,long2016coordinated}. It seems therefore promising to build on the experience of historical apps to improve new apps. Additionally, prior work~\cite{li2019cocotest,bevan2002guidelines} have demonstrated that interactions among developers of different software can be effective to improve the quality of each software.

{\bf This paper.} We hypothesize that if app \textbf{A} and app \textbf{B} belong to the same category (we consider the categories listed in the Wikipedia enumeration of popular free and open source Android apps\footnote{\url{https://en.wikipedia.org/wiki/List_of_free_and_open- source_Android_applications}}, e.g., web browsers, Games, etc.), bug reports from one can be relevant for discovering bugs in the other. Unfortunately, there can be too many bug reports filed in some app categories. For example, in the web browser category, the issue tracker of Firefox alone has received over 20 thousand bug reports. It is therefore necessary to identify those issues that are more likely to be relevant for the app under-study (i.e., the target app for bug discovery). To that end, our novel strategy in this work is to explore app reviews written by users for the target app. Our idea is that app reviews, which may contain hints about buggy behaviour observed by users of the target app, can be matched to bug reports from other apps in the same category.

\begin{figure}[!t]  

	\centering
	\subfigure[Bug report of Signal.]{
		\includegraphics[width=5cm]{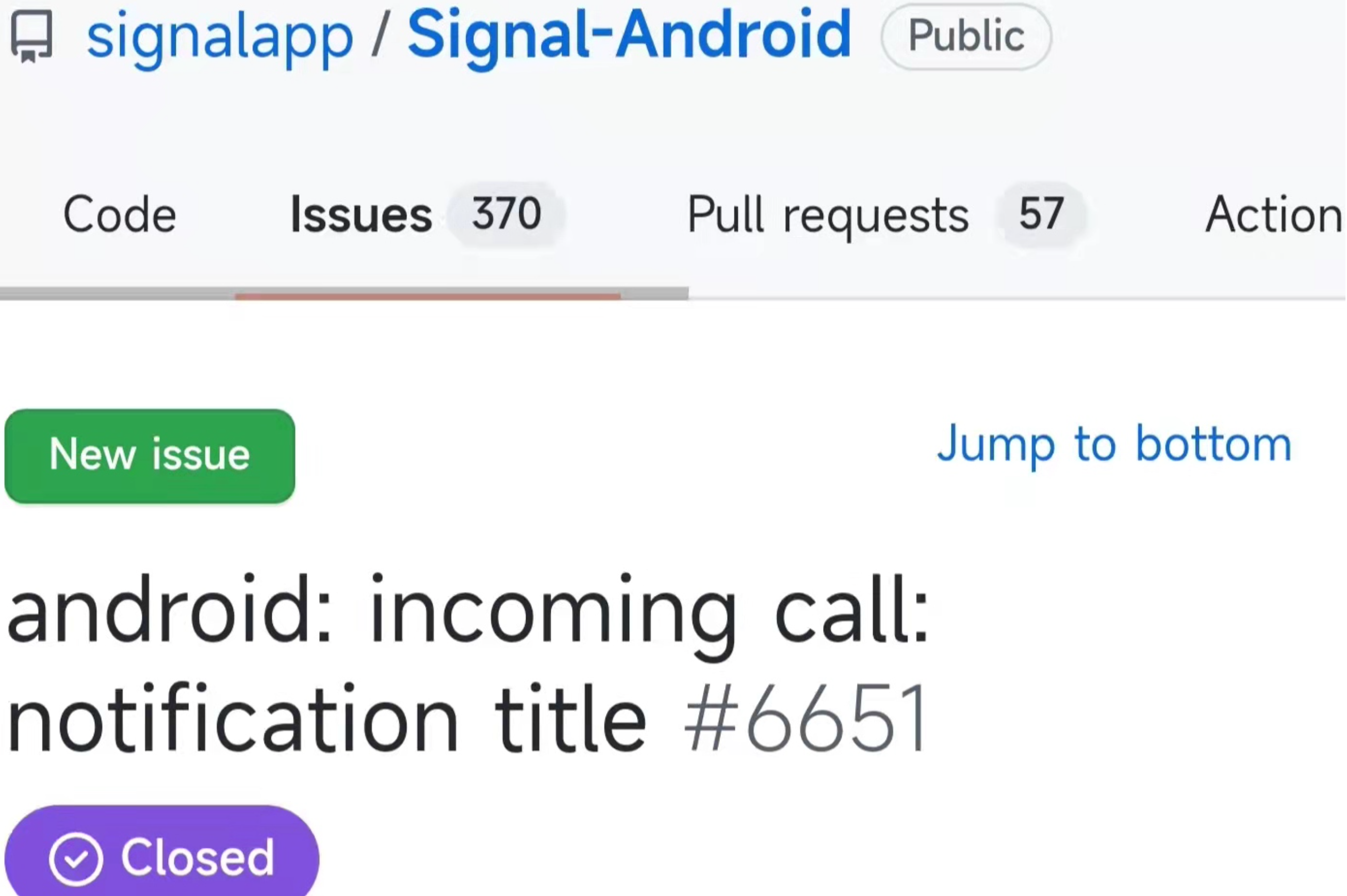}
		\label{fig: sdrport}
	}
	\subfigure[App review of Wire.]{
		\includegraphics[width=5cm]{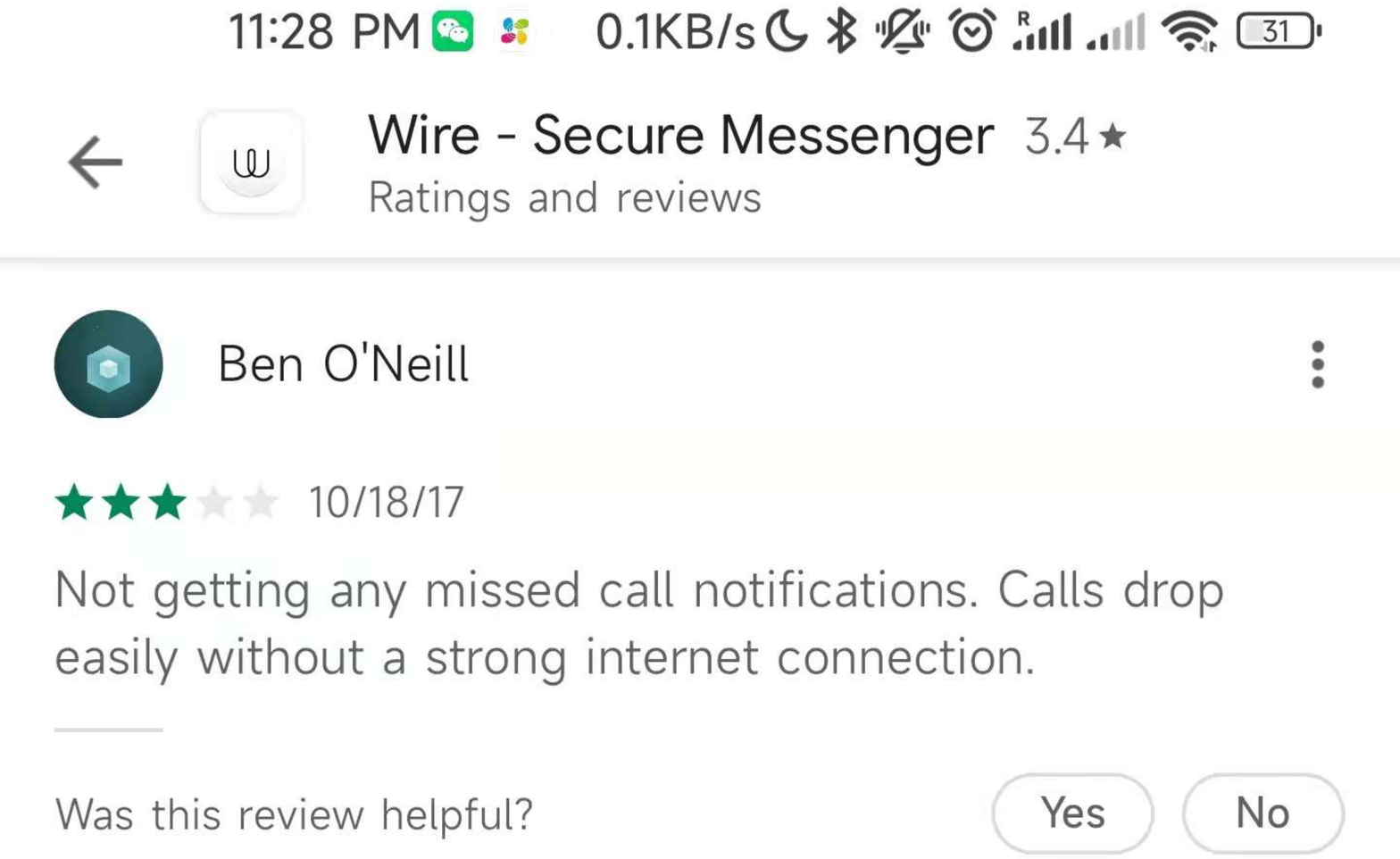}
		\label{fig: matchreview}
	}
	\caption{Example of bug report and app review matched by \tool.}
	\label{fig: sdreport_matchreview}	
\end{figure}

\begin{figure}[t!]  

	\centering
	\subfigure[Newly opened issue for Wire.]{
		\includegraphics[width=5cm]{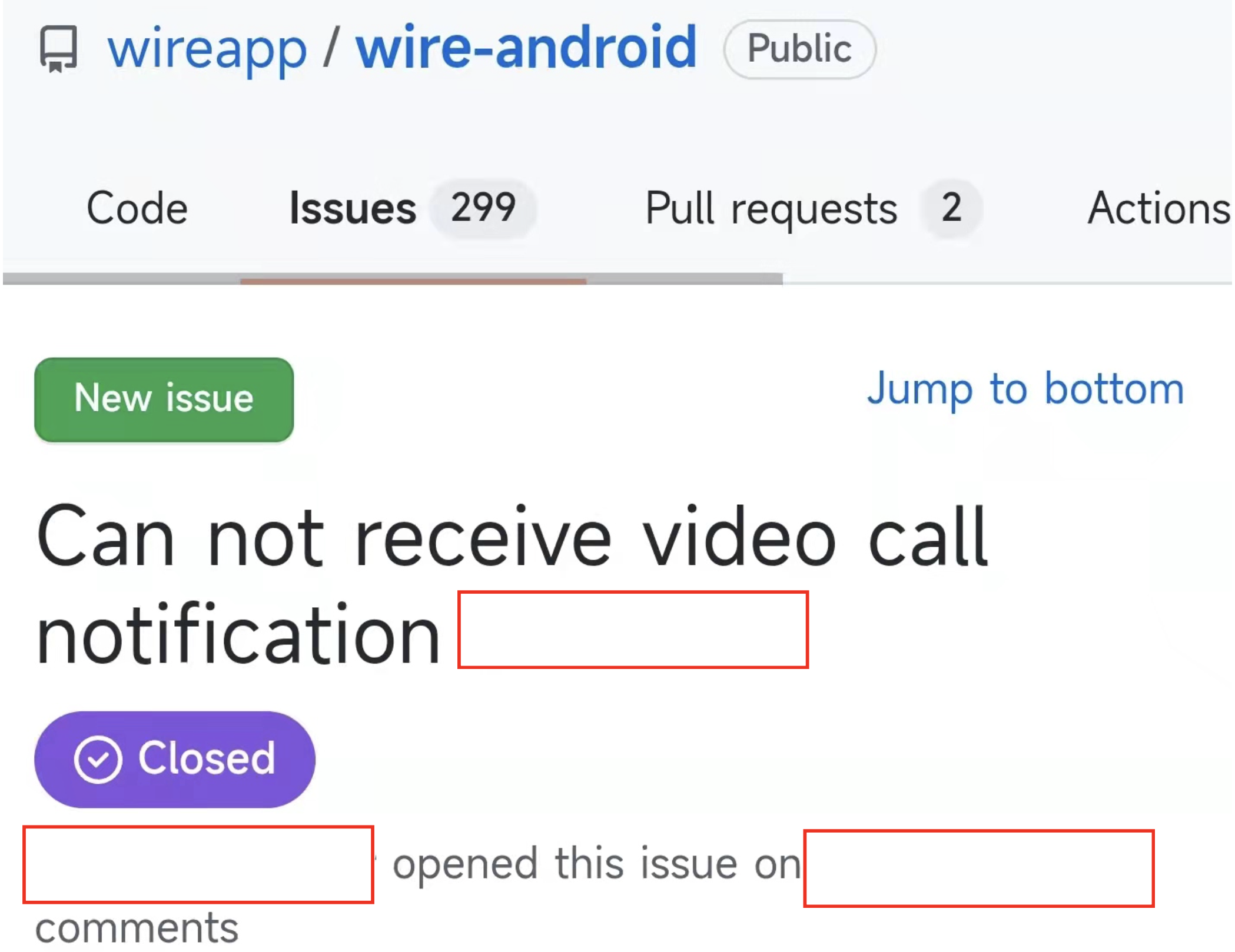}
		\label{fig: issue_B}	
	}
	\subfigure[Response from official developer.]{
		\includegraphics[width=5cm]{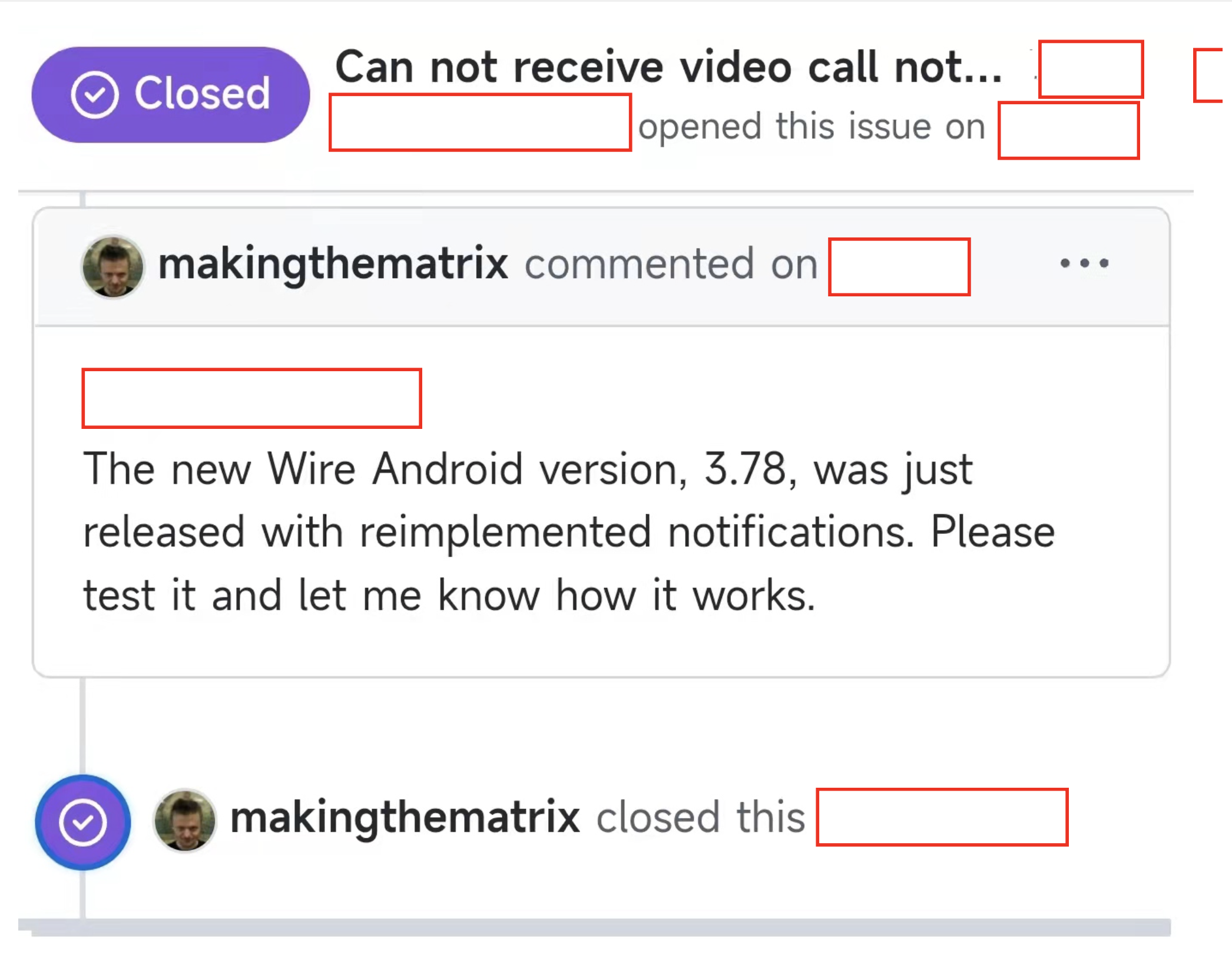}
		\label{fig: response_B}	
	}
	 \caption{Bug found by \tool and response from official developer (some texts have been hidden for privacy protection).}
	 \label{fig-bug}
\end{figure}

We propose \tool, a collaborative bug finding approach that is guided by user app reviews. \tool finds bugs by recommending a bug report of app {\bf A} (e.g., the excerpted bug report of Signal in Figure~\ref{fig: sdrport}) as being relevant to the target app {\bf B} (i.e., Wire) given the similarity of the bug report with app reviews from users of {\bf B} (e.g., the excerpted user review of Wire in Figure~\ref{fig: matchreview}). 
With the app review in {\bf B} matched with a similar bug report in {\bf A}, we reproduce the bug in {\bf B} by leveraging reproduction steps in the bug report and additional information details from the app review.
If reproduction is successful, we can confirm having found a ``new bug'' that will be filed into the official issue tracker for app \textbf{B}.
For example, the corresponding bug found by \tool in Wire was reported to its developers (shown in Figure~\ref{fig: issue_B}), and was finally got fixed by Wire's official developer (Figure~\ref{fig: response_B}) in one day.
Surprisingly, the relevant user app review had been submitted since four years, but there is no any corresponding bug reported in the official issue tracker of the app yet. 
To the best of knowledge, this scheme of collaborative bug finding driven by app reviews, has not been previously explored in the literature. 



The main contributions of our work are as follows:

\begin{itemize}[leftmargin=*]
    \item We present insights from an empirical study about the similarity of issues reported across apps from the same category. These insights provide the intuitive basis for collaborative bug finding. 
    \item We devise \tool{}, a bug recommendation system, which leverages similarities of user app reviews with bug reports to identify which bug reports from same category apps are good candidates to attempt a bug reproduction on a target app.   
    \item We demonstrate experimentally the effectiveness and usefulness of \tool{}: applied to 6 apps from different categories, we find and reproduce 20 new bugs (i.e., bugs that are not yet reported in the issue trackers of these apps): 9, 6, and 4 reports have been already triaged, confirmed, fixed by official development teams, respectively.
\end{itemize}




%% file: 2.Motivating.tex
\section{Preliminary Study}
\label{sec:motivating}

In this section, we conduct a preliminary study to evaluate the hypothesis of our work. Specifically, we seek to check that apps from the same category are more likely to share similar bug reports (and thus a bug report from app A could be relevant for app B if A and B are from the same category). We focus the comparison by estimating the overlap (i.e., the proportion of common words) between bug reports. 
To that end, we analyze the overlap rate of top$_K$ frequent words of reports for apps from same or different categories. From a more qualitative perspective, we also analyze which types of words are frequently shared.

\begin{figure*}[htbp!]  
	\centering
	\includegraphics[width=1\linewidth]{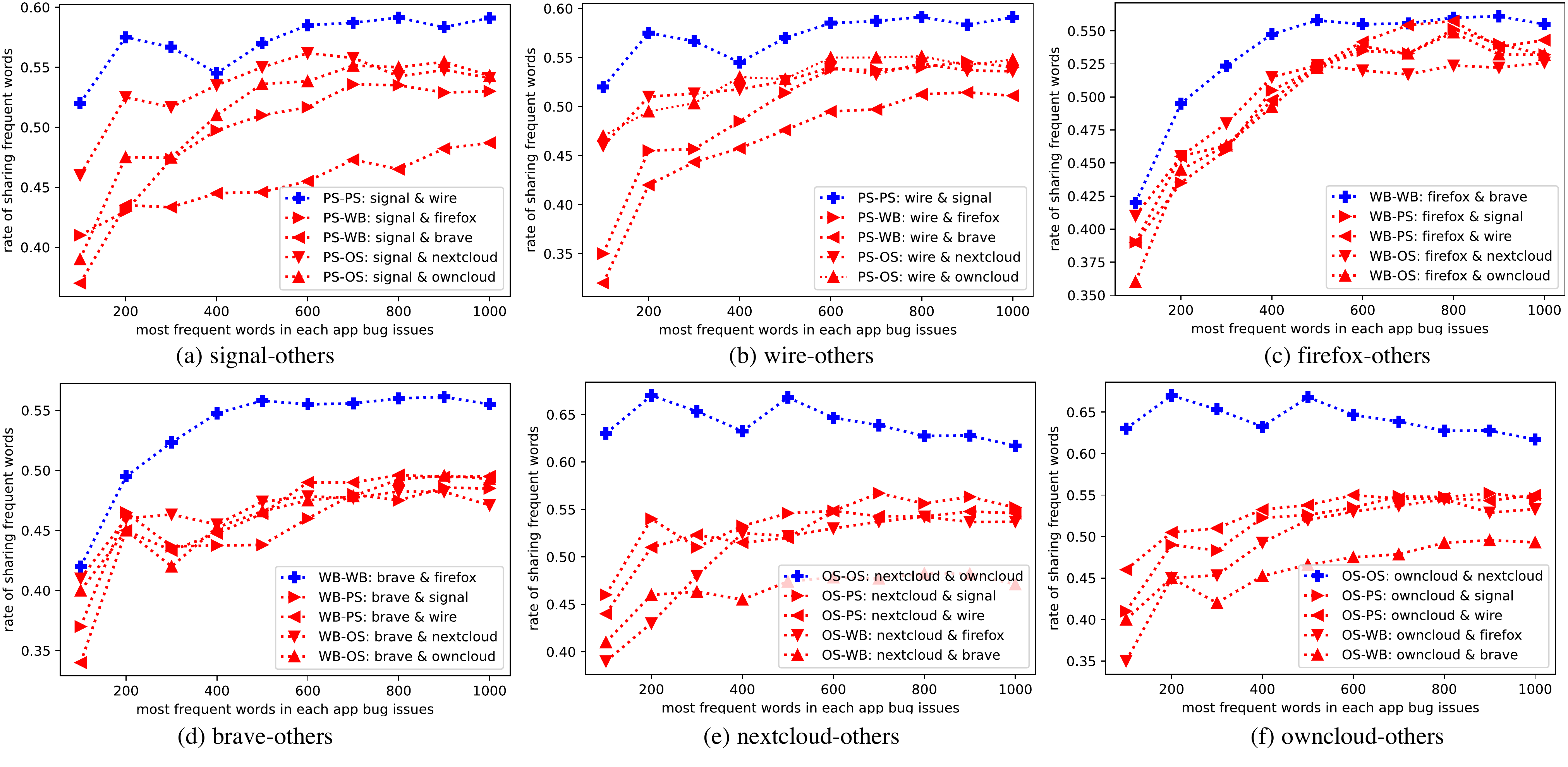}
	
	\caption{Overlap rate of hot words of category apps' bug issues.}
	\label{Fig: RQ1a}
\end{figure*}

\subsection{Empirical Setup}

\textbf{Apps: }
To conduct our preliminary study, we consider six popular apps listed in the Wikipedia page of free and open-source Android apps. Table~\ref{tb: Apps} summarizes statistics about these apps. We consider \textit{Signal} and \textit{Wire} within the \textit{Privacy-Security} category. Both apps have been downloaded more than one million times from the Google Play store. We also consider \textit{Firefox} and \textit{Brave}, two widely popular in the \textit{Web Browser}, category. 
Finally, we consider \textit{Nextcloud} and \textit{Owncloud} among the apps in the \textit{Office Suite} category. 

\begin{table*}[!t]
\caption{The Apps and their category.}
\resizebox{0.96\textwidth}{!}{
\begin{tabular}{lllrr}
\toprule
{\bf App Name}        & {\bf Repo Name}      & {\bf Category}         & {\bf\# Downloads}               & {\bf\#Bug Reports}  \\
 \midrule
 Signal          & Signal-Android & privacy security & \textgreater 50 million   & 11,980                      \\ \hline

Wire       & wire-android & privacy security & \textgreater 1 million   & 3,677   \\ \hline

Firefox Android & fenix          & Web Browser      & \textgreater{}100 million & 24,087             \\ \hline

Brave           & brave-browser  & Web Browser      & \textgreater{}50 million  & 21,436            \\ \hline
 
NextCloud  & nextcloud    & Office Suite     & \textgreater 1 million   & 9,890  \\ \hline
 
owncloud   & owncloud     & Office Suite     & \textgreater 0.1 million & 3,571    \\ 
\bottomrule

\end{tabular}
}
\label{tb: Apps}
\end{table*}

\textbf{Bug Reports: }
For each app, we have collected all bug reports that are present in their issue tracker system. The number of collected bug reports is given in the last column of Table~\ref{tb: Apps}. 
Note that, for our experiment, we employ the Python NLTK~\cite{loper2002nltk} library and self-defined filters to pre-process the bug reports for natural language processing. We apply typical pre-processing tasks to remove stop words~\cite{wilbur1992automatic}, punctuation, digits, etc.\cite{haddi2013role}. 
Meanwhile, to limit experimental bias, we set 10 different sizes $K$ for the set of most frequent words, increasing step-wise until an order of magnitude: we consider Top$_{100}$, Top$_{200}$,..., Top$_{1000}$ frequent words. 
Concretely, to build each Top$_K$ set for each app, we extract the $K$ most frequent words in its bug reports. 
By analysing the Top$_K$ frequent words, we can assess the differences in shared words between bug reports of same-category apps and different-category apps. Applied to all set ten Top$_K$ sets, we can further check for potential trends, while empirically identifying the value of $K$ under which the overlap (i.e., the proportion of shared words) is the highest. 


\textbf{Notation: }
In the rest of this paper, the category Privacy-Security, the category Web Browser, and the category Office-Suite are referred to as \texttt{PS}, \texttt{WB}, and \texttt{OS}, respectively. 
We also note the three pairwise combination of apps from the same category as follows: 
\texttt{PS-PS}: <Signal, Wire>; \texttt{WB-WB}: <Firefox, Brave>; \texttt{OS-OS}: <Nexcloud, Owncloud>.
Similarly, we consider 12 pairwise combinations of apps from different categories as follows: 
\texttt{PS-WB}: \texttt{<}Signal, Firefox\texttt{>}, \texttt{<}Signal, Brave\texttt{>}, \texttt{<}Wire, Firefox\texttt{>}, and \texttt{<}Wire, Brave\texttt{>}; \texttt{PS-OS}: \texttt{<}Signal, Nextcloud\texttt{>}, \texttt{<}Signal, Owncloud\texttt{>}, \texttt{<}Wire, Nextcloud\texttt{>}, and \texttt{<}Wire, Owncloud\texttt{>}; \texttt{WB-OS}: \texttt{<}Firefox, Nextcloud\texttt{>}, \texttt{<}Firefox, Owncloud\texttt{>}, \texttt{<}Brave, Nextcloud\texttt{>}, and \texttt{<}Firefox, Owncloud\texttt{>}.

\subsection{Hypothesis Analysis} \label{sec: general}

In this section we check whether our hypothesis is valid from the quantitative and qualitative aspects of bug reports.

{\bf Quantitative Analysis:}
The results of the quantitative study are presented in Figure \ref{Fig: RQ1a}, where for each app X, we compute the percentage (or rate) of overlap, i.e., the percentage of shared words between the Top$_K$ frequent words in the bug reports of X and the Top$_K$ frequent words in the bug reports of another app Y. 
We note that, except for the \texttt{<}Owncloud, Nextcloud\texttt{>} pair, {\bf both same-category and different-category pairs present that the percentage of shared words increases or reaches a plateau (often after K=500)}.
In all six diagrams, and for all  $K \in \{100,200,...,1000\}$, the percentage of shared Top$_K$ frequent words is the highest when apps X and Y are from the same category (i.e., for Signal and Wire, Firefox and Brave, and, Owncloud and Nextcloud). 
It indicates that, {\bf bug reports of same-category apps could present higher similarity than bug reports of different-category apps.}

\begin{figure*}[htbp!]  
	\centering
	
	\includegraphics[width=1\linewidth]{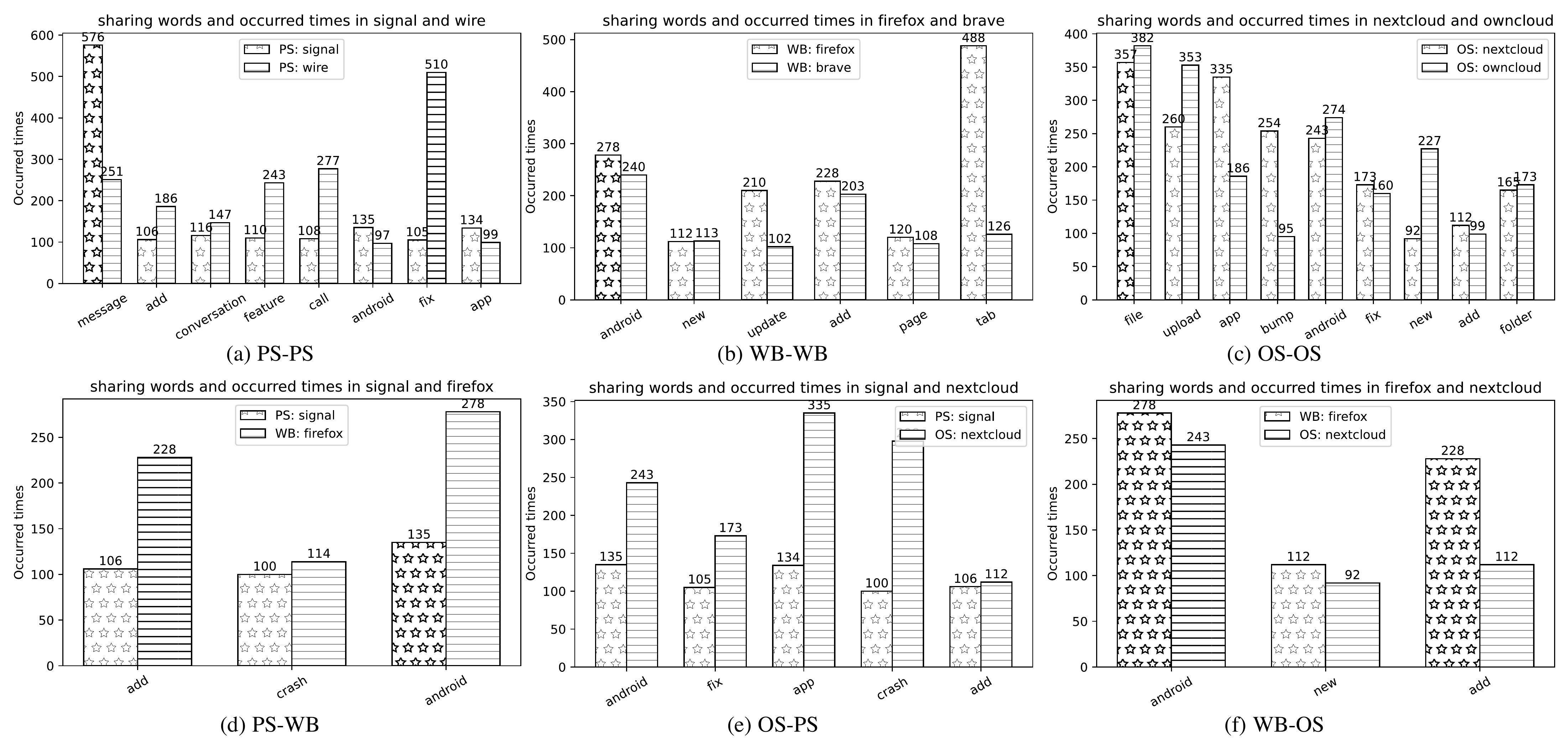}

	\caption{Distribution of Top$_{20}$ shared frequent words in same-category app pairs and different-category app pairs.}
	\label{fig: RQ1b}
\end{figure*}

{\bf Qualitative Analysis:}
We further investigate to what extent frequent words in bug reports of same-category and different-category apps share. 
To this end, we conduct the quantitative analysis by setting the threshold of frequency as 20 to select the most frequent words in bug reports (all words are trimmed to keep their trunk, e.g., crashes$\rightarrow$crash).
With the top-20 most frequent words of each app, we assess to what extent different app pairs share these words, and how frequent the shared words occur in each app. 
Table \ref{tb: freq20} lists the top$_{20}$ frequent words of each app.

\begin{table}[!t]
	\centering
	\setlength\tabcolsep{2pt}
\caption{Top$_{20}$ frequent words in bug reports for each app.}
\resizebox{0.9\textwidth}{!}{
\begin{tabular}{c|c|L{65mm}}
\toprule
\multicolumn{2}{c|}{\bf App} & {\bf Top$_{20}$ frequent words} \\ \midrule
\multirow[b]{2}{*}{PS} & Signal &
  signal, message, sms, group, android, app, contact, send, mms, conversation, notification, feature, call, add, fix, request, crash, textsecure, text\\ \cline{2-3}
& Wire &
  fix, feature, add, conversation, avs, part, message, bump, wire, new, update, call, user, remove, app, android, video, version \\ \hline
\multirow[b]{2}{*}{WB}& Firefox &
  bug, fnx, android, tab, add, tabs, update, search, issue, menu, components, fenix, button, page, crash, new, ui, strings, version\\ \cline{2-3}
& Brave &
  brave, x, release, android, chromium, add, desktop, test, run, rewards, manual, ads, tab, browser, upgrade, wallet, new, settings, page, update \\ \hline
\multirow[b]{2}{*}{OS}& Nextcloud &
  app, upload, bump, android, crash, file, nextcloud, fix, folder, auto, stable, add, error, rc, new, use, account, server\\ \cline{2-3}
& Owncloud &
  android, upload, new, file, app, folder, fix, owncloud, feature, release, bug, request, update, share, add, view, arch, bump \\ \bottomrule
\end{tabular}}
\label{tb: freq20}
\end{table}

We have three pairs of same-category apps and twelve different-category apps. 
Considering the space limitation, we only show three pairs of different-category apps, and all data of twelve pairs of different-category apps are publicly available in our \href{https://github.com/BugRMSys/BugRMSys/blob/main/12_different_categories_apps}{replication package}\footnote{\url{https://github.com/BugRMSys/BugRMSys/blob/main/12\_different\_category\_apps}}.
For the \textbf{PS} category, 
as shown in Figure \ref{fig: RQ1b}(a), the top frequent words shared by Signal and Wire include message, conversation, call, etc. 
Apparently, these shared words represent the key features of these two apps (i.e., communication tool). 
They also share some general hot words about the Android Framework, such as android and app. 
For the \textbf{WB} category (cf. Figure \ref{fig: RQ1b}(b)), the words shared by Firefox and Brave contain page and tab, of which both are highly relevant to the web browser. 
While for the category of \textbf{OS} (cf. Figure \ref{fig: RQ1b}(c)), Nextcloud share ``fix, file, folder, upload'' words with Owncloud, which are related to the cloud storage system.
Such results indicate that 
{\bf the bug reports of same-category apps indeed share their field-specific information}.
On the other hand, as shown in Figures \ref{fig: RQ1b}(d), \ref{fig: RQ1b}(e), and \ref{fig: RQ1b}(f), the bug reports of different-category app pairs share less words than the same-category app pairs,
and the shared words are mainly about the 
general information or problem of the Android framework itself (with words such as android, app, crash, add).

Regarding words that are not shared, by differentiating the top$_{20}$ frequent words and the shared words of the same/different-category pairs, we note that those non-shared frequent words of the same-category app pairs are about general features or problems (e.g., crash, notification, etc.), while those non-shared frequent words of the different-category app pairs are related to their main features of those apps. 
For example, given <Signal, Firefox>, a different-category app pair, the non-shared words in Firefox contain tabs, browser, pages, ads, which are associated with the traditional features of web browser: advertisement, web pages, etc.

To sum up, {the bug reports of apps in the same category  share more similar information than the apps in different categories, and the shared information is prone to the specific features related to the category with the frequently used words}.


\find{{\bf Hypothesis} \ding{224} \ding{125}With the commonness/difference on the quantitative and qualitative analysis of bug reports in the same/different categories, we would like to intuitively hypothesize that it would be feasible to employ experience (e.g., bug reports) from same-category apps and the community (reviews) of target app \textit{B} to discover potential bugs of \textit{B}.\ding{126}}

%% file: 3.Approach.tex
\section{\tool{}}\label{sec:methodology}
\begin{figure}[!ht]
    \centering
    \includegraphics[width=1\linewidth]{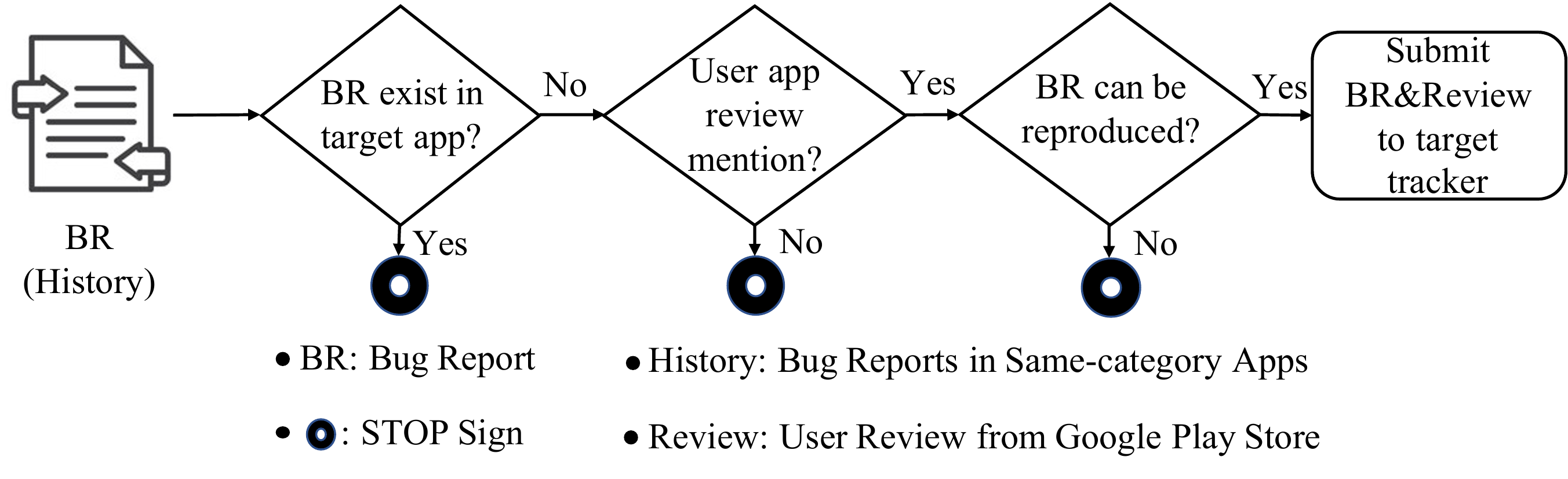}
    \caption{General workflow of app review driven collaborative bug finding.}
    \label{fig: generalworkflow}
\end{figure}

Figure~\ref{fig: generalworkflow} depicts the general workflow of the approach for app review driven collaborative bug finding.\\
\noindent
{\bf\em Notation:} we will refer to app \texttt{A} and app \texttt{B} as two apps that belong to the same category.

\begin{figure*}[!h]
    \centering
    \includegraphics[width=1\linewidth]{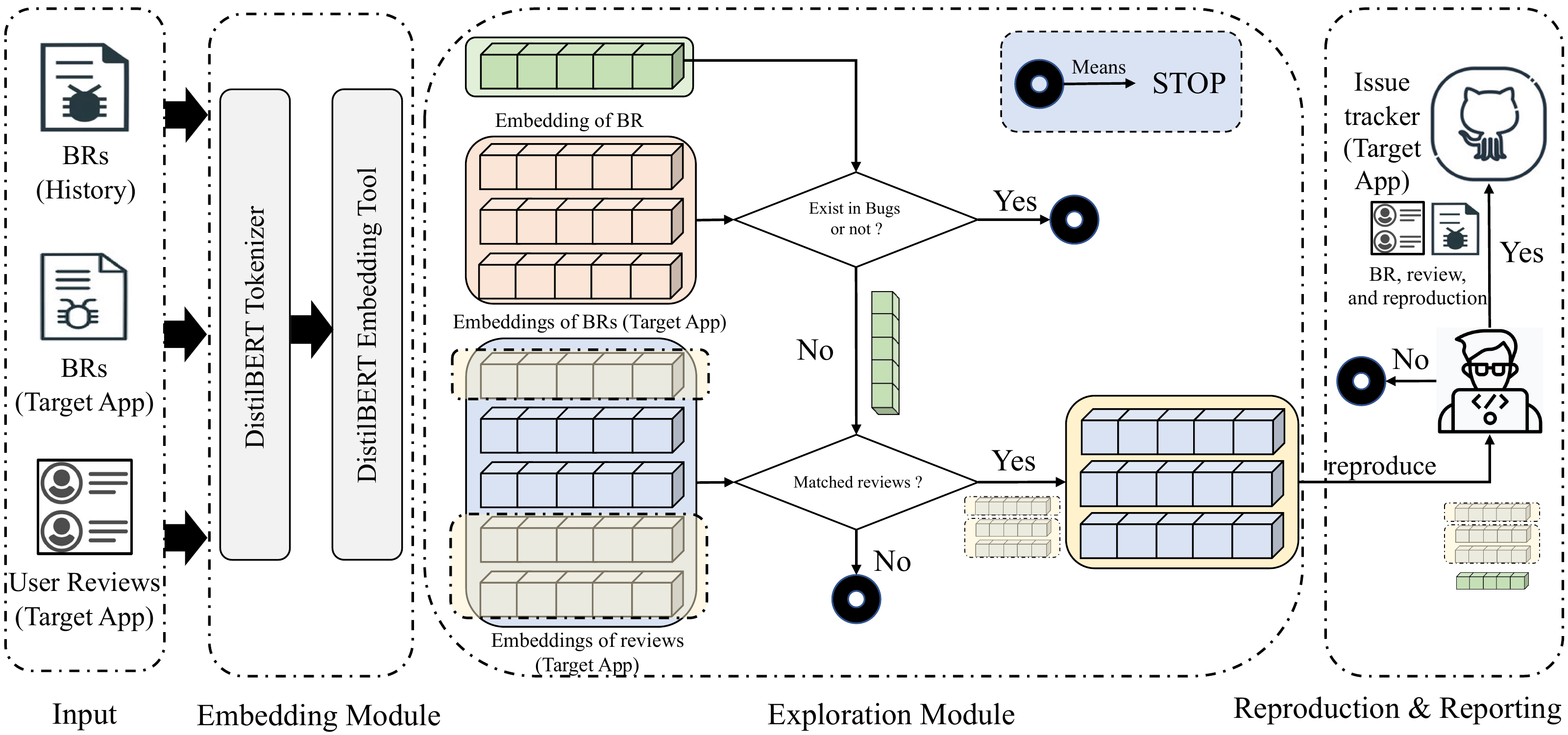}
    \caption{Overview of \tool{}.}
    \label{fig: technicalworkflow}
\end{figure*}

Our approach attempts to leverage the development experience of the historical app {\tt A} to find bugs in the target app {\tt B}. The workflow therefore starts with a representative bug report that has been handled in the development of app {\tt A}. If a similar bug report exists in the issue tracker of the target app {\tt B} under study, the bug finding process is halted and must be restarted with another bug report from app {\tt A}. Otherwise, the workflow proceeds to check whether the bug report content is similar to some user reviews (that has therefore gone overlooked). If one or several relevant app reviews are found, we must attempt to reproduce the bug in app {\tt B} based on reproduction steps in the bug report of app {\tt A} as well as specific details in user review of {\tt B}. In our evaluation, once the buggy behavior is confirmed through reproduction of the bug report, we further submit a new bug report in the issue tracker of app {\tt B}.
In the remainder of this section, we will present a real-world example before detailing the technical approach for bug recommendation.

\subsection{Running Example} \label{sec: re}
\begin{figure}[!t]
    \centering
    \includegraphics[width=1\linewidth]{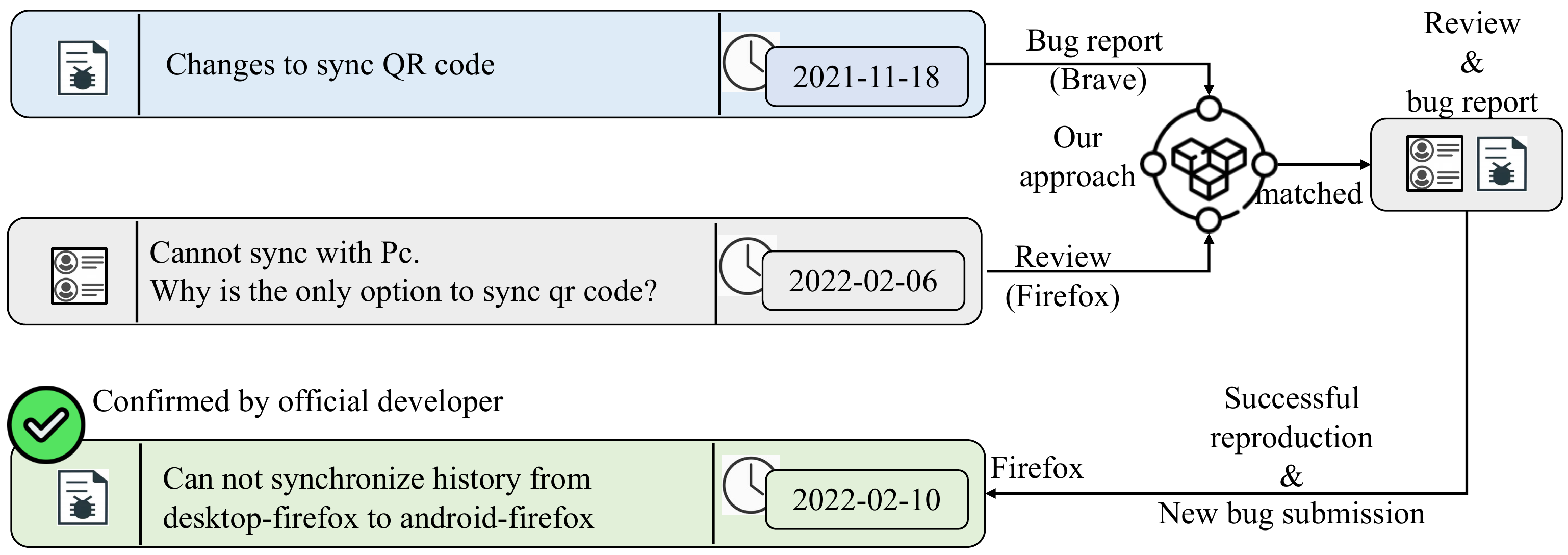}
    \caption{Running example of \tool{}.}
    \label{fig: app_review}
\end{figure}

Figure~\ref{fig: app_review} illustrates the case where we leveraged \tool{} to discover a new bug in the web browser app Firefox. By iterating over bug reports from the active development repository of Brave, we identified a bug report which refers to synchronization with QR Code. A similar bug report was absent from the issue tracker of Firefox. After matching by \tool{}, a user review of Firefox had clearly stated a similar problem ``{\em Cannot sync with Pc. Why is the only option to sync qr code?}''. With the user's assessment, the bug report might indeed be relevant, we thus explore the steps enumerated in Brave's bug report
to reproduce the matched problem in Firefox. 
Surprisingly, the bug was successfully reproduced within a few minutes. We then submitted a bug report with two screenshots into the issue tracker of Firefox, which was eventually confirmed by the Firefox development team in 4 days.

Intuitively, our bug recommendation could have started with considering available user app reviews and try to correlate with historical bug reports from other apps in the same category of the target app. Unfortunately, in practice, most app reviews do not provide usable information. Therefore, we propose to initiate the search with the bug reports, which are in lesser numbers, and are more structured. 
Nevertheless, many bug reports in the same category as the target app are actually raising irrelevant issues. Therefore, it is important to further check if such potential issues have caught user attention and lead them to write reviews that mention them. This motivates the need to devise a reliable mechanism to precisely match useful app reviews with relevant bug reports.

\subsection{Approach}
Figure \ref{fig: technicalworkflow} illustrates the details of automatically finding bugs with \tool{}, which includes two main modules: an embedding module and an exploration module. The remainder of this section describes how the embedding module deals with bug report and app review text representation, and how the exploration module eventually identifies relevant bug reports for recommendation.

\subsubsection{\bf Embedding} \label{sec: pretrain}
To embed app reviews and bug reports into vector representation for natural language text semantic similarity computation, we resort to state-of-the-art pre-trained deep learning models. 
In practice, when reasoning about app reviews and bug reports, we face at least two major challenges: 

\begin{enumerate}[leftmargin=*]
    \item App reviews and bug reports are organized differently, and their vocabularies differ from each other (technical vs non-technical terms). In addition, app reviews are more error-prone (e.g., spelling mistakes), may include emoticons, and may contain non-technical references (e.g., swearing). Reviews also often repeat, i.e., the same review can be submitted by the same user several times. 
   \item Bug reports are written based on a technical template that includes description, title, reproduction steps. They can also include code-specific technical terms such as test names (e.g., ``{\tt RewardsBrowserTest}''). A complete bug report is often very long, hence difficult to capture overall semantics. 
\end{enumerate}

Towards providing a precise and effective representation model for bug reports and app reviews, we address each challenge specifically in \tool{}:

For \textbf{challenge (1)}, 
we use common NLP pre-processing techniques to eliminate noisy words and/or reviews. We first lowercase all tokens and remove punctuation as well as stop words~\cite{panichella2015can,villarroel2016release,stanik2018simple}. We then use a regularization function to remove digits and emoticons. In addition, we drop too short and too long reviews based on outlier thresholds: empirically, we determined that reviews containing fewer than 10 words or more than 200 words should be dropped as they are outliers. We adopt the rule-based methods in ~\cite{man2016experience,vu2015mining} to correct word repetitions and spelling mistakes. 

For \textbf{challenge (2)}, we must carefully process bug reports to extract relevant information. Indeed, prior work~\cite{zhou2016combining,martens2019extracting} have established that the detailed bug report descriptions usually contain significant noise, including the report template itself. Previous study stated that the title of a bug report already includes the essential content of the bug report description~\cite{lamkanfi2010predicting,ko2006linguistic,wang2008approach}. Thus, in our approach, we consider the bug report titles as input to drive the recommendation system. Other details (e.g., reproduction steps) will be considered in the reproduction and submission phase. 

To address the challenge for capturing semantic features from the app reviews and bug reports, we resort to an effective method of converting bug reports and app reviews into numerical representations. There are various techniques~\cite{zhang2010understanding,manning1999foundations,pennington2014glove,mikolov2013efficient,joulin2016bag} in the NLP community for mapping texts into tokens and low-dimension representations. 
In recent years, pre-trained deep learning methods~\cite{devlin2018bert,sanh2019distilbert} have become popular across communities given their practicality and their effectiveness. 
In this work, we consider DistilBERT~\cite{sanh2019distilbert}, a pre-trained lightweight network model built on the top of BERT~\cite{devlin2018bert}, that offers similar performance with transformer-based models as BERT,
but requires significantly less computer resource and training/tuning time than BERT. 
And DistilBERT employs a byte pair encoding method~\cite{sennrich2015neural} as its word tokenizer which empowers it to alleviate ``out of vocabulary'' issues and to recognize compound words, especially for variables and camel-case (and other composition-based) named elements. 
We reuse the same function defined in~\cite{devlin2018bert} and apply its default setting to get the embeddings of both bug reports and app reviews.
The embeddings of bug reports and app reviews can be used for calculating the document's cosine similarity.


\subsubsection{\bf Exploration} \label{sec: newb}
Once the embedding module has produced numeric representations for bug reports and app reviews, the exploration module attempts to identify those reviews that are relevant for a given bug report. If a user review is found, the bug report can be recommended (supposing that no similar bug report was already filed for the target app). 
To this end, following empirical findings from previous studies~\cite{ghosh2006similarity} dealing with text similarity metrics~\cite{wooditch2021getting}, we resort to employing the cosine similarity~\cite{sitikhu2019comparison} between low-dimensional representations of bug reports and app reviews to measure their relatedness. 
\tool{} then prioritizes app reviews as being potentially relevant or not by referencing a threshold value to dismiss app reviews that should not be listed. 
Therefore, if one app review of B is found to be similar with a bug report in A, the bug report will be recommended as describing a bug that is relevant for the target app B.
Otherwise, the bug report will not be recommended. 
The threshold setting is presented in Section~\ref{sec:rq1}.



%% file: 4.Emperical_design.tex
\section{Experimental Design}\label{sec:exp_design}


\subsection{Research Questions}
\label{sec:rqs}


Our experiments aim at answering the following research questions:
\begin{itemize}[leftmargin=*]
    \item {\bf RQ-1: To what extent is the \tool{} bug recommendation approach driven by app reviews effective?}
    Our approach is devised to recommend bugs with our hypothesis. In this RQ, we first investigate to what extent app reviews can be used to recommend the bug report for the target app. Before that, we build a ``ground truth'' set of bug reports pairwise combinations (from a target app and a same category app).

\item \textbf{RQ-2:} \textbf{Can \tool{} expose bugs in real-world apps based on existing app reviews?}
With this research question, we explore the feasibility of exposing bugs in real-world apps by \tool{} with 20 apps of 9 categories.
We investigate both the Exploration module for bug recommendations, as well as the actual bugs that developers can confirm.


\item \textbf{RQ-3:} \textbf{How does \tool{} compare against previous bug recommendation systems?}
With this research question, we compare the bug recommendation approach of \tool{} against the bug recommendation systems (DeepMatcher~\cite{haering2021automatically} and Bugine~\cite{tan2020collaborative}) in three dimensions (inputs, automation, effectiveness). 

\end{itemize}

\subsection{Dataset}
\label{sec:data}

We investigate the feasibility of our hypothesis with \tool{}, we first curate a dataset that collects bug reports and app reviews from 20 free and open source apps of 9 categories. For each app, we collect bug reports from its GitHub issue tracking system. 
App reviews are selected by checking whether an app review is related to a concrete bug report.
App reviews are further curated following the order given after a sort on ``helpful''.
A review is given a ``helpful'' score according to the number of users who agree with it. In a recent study study, H{\"a}ring et al.~\cite{haering2021automatically} have used this score to measure the importance of a review. 
Note that, collected bug reports and app reviews are written in English. 
Table~\ref{tab:all-data-category} summarizes the information of collected data.

\begin{table}[!h]
\caption{Dataset Summary.}
\resizebox{\linewidth}{!}{
\begin{tabular}{lcrr}
\toprule
{\bf App Category}   &   {\bf \# Apps}  &   {\bf \# bug reports}   &  {\bf \# app reviews}   \\ \midrule

privacy security & 2 & 15,657    & 15,428 \\ \hline
web browser& 2 & 45,523    & 166,360       \\  \hline
office suites &2& 13461    & 2,260      \\  \hline
emulator & 3 &25,822    &  43,709    \\ \hline
communication & 2&4,555    & 3,166    \\ \hline
game & 2&1,590    & 3,098     \\ \hline
multimedia & 2&5,898    & 53,746  \\ \hline
reading &2 & 3,704  & 62,755 \\ \hline


science and education &3 & 16,045   & 36,413\\ 
  \midrule

{\bf Total}     & 20 & 132,255  & 386,935   \\ \bottomrule

\end{tabular}}
\label{tab:all-data-category}
\end{table}

\subsection{Evaluation Metrics}
\label{sec:metrics}
Our study leverages a variety of metrics to validate the experiments.
First, we rely on overlap rate analysis~\cite{kowalski2010information} to investigate our hypothesis about the similarity of bug reports across same-category apps. 

\textbf{Overlap Rate:} Given two sets $X$ and $Y$, Overlap rate of $X$ with $Y$ is computed as follows:
\begin{equation}
    \begin{split}
        Overlap(X_Y) = \frac{size(\left| X \cap Y \right|)}{size(X)},
    \end{split}
\label{eq: overlap}
\end{equation}
where $size$ denotes the size function for sets. 
If both sets have the same size (e.g., in our case, we select top$_K$ frequent words in the different sets of bug reports), 
then $Overlap(X_Y)$ == $Overlap(Y_X)$. 

To evaluate the overall performance of \tool{}, we use Acc@N and Mean Recriprocal Rank (\textit{MRR}), 
which are widely used  metrics for recommender systems~\cite{ye2014learning,zhou2012should,poerner-etal-2020-e}.

\textbf{Acc@N:} 
Acc@N hit measures the retrieval precision over the top$_N$ recommended issues or reviews in the ranked list:
\begin{equation}
    \begin{split}
        Acc@N = \frac{ \Sigma^N_i(pair(i))}{LENGTH},
    \end{split}
\label{eq: acc@N}
\end{equation}
where \textit{LENGTH} represents the length of tested ground truth; pair(i) means if the \textit{i-th} issue \textit{B} hit the \textit{i-th} top$_N$ reviews relevant to issue \textit{A}, if yes, \textit{pair(i)} = 1, else 0. Overall, \textit{Acc@N} is an approach used in previous research describing how often the issue in target \textit{B} is among the top$_N$ Nearest Neighbours (by cosine) of an DistilBERT word space.

\textbf{MRR:} \textit{MRR} is short for mean reciprocal rank and is a popular metric used to evaluate the efficiency of recommendation system \cite{shani2011evaluating,cames2006recommendation,mahmood2009improving}. The equation of accuracy is described as follows:
\begin{equation}
    \begin{split}
        MRR = \frac{1}{N}*\Sigma_{i=1}^{\left|N\right|}\frac{1}{rank_i},
    \end{split}
\label{eq: mrr}
\end{equation}
where \textit{N} is the length of ground truth; For the \textit{i-th} issue (app A) in ground truth, rank$_i$ represents the position of recommended review which also relevant to issue in app B.


\subsection{Implementation \& Availability}
\label{sec:implementation}
We first developed two crawlers for automatically collecting bug reports and app reviews based on python packages: PyGithub\footnote{\url{https://pypi.org/project/PyGithub/}} and google play scraper\footnote{\url{https://pypi.org/project/google-play-scraper/}}, respectively. To filter most useless reviews, we only select reviews in the text size from 10 to 200.
We have implemented a prototype version of \tool{} using Python (and associated frameworks) with a well-known, light-weight, transformer-based, and contextual pre-trained model, DistilBERT, to extract vector representations for both bug reports and app reviews. 
The dataset and the replication package of \tool{} are publicly available at: \begin{center}
    {\bf\url{https://github.com/BugRMSys/BugRMSys}}
\end{center} 

%% file: 5.Emperiment_results.tex
\section{Experimental Results}\label{sec:evaluation}
In this section, we conduct qualitative and quantitative analysis to evaluate \tool{}. To this end, we evaluate the effectiveness of \tool{}, we compare \tool{} with state-of-the-art tools for bug recommendation, we study the characteristics of \tool{}, and we explore the transferability of \tool{}.

\subsection{[RQ-1]: Effectiveness of \tool{}}
\label{sec:rq1}

To answer RQ1, due to the huge manual effort for assessing similarity of app reviews with bug reports, we focus on a single same-category app pair (FireFox, Brave),
where FireFox will be the input app and Brave the target app (i.e., we use bug reports from Firefox to find new bug reports in Brave by matching Brave user reviews). Note that in the running example (cf. Section~\ref{sec:motivating}), we had illustrated a bug case where Firefox was the target app and Brave was the input app. This is an additional argument that \tool{} can explore the experience of any app and leverage it for any other same-category counterpart regardless of which app appears to have more historical data.

We start by building a ground truth dataset to assess the ability of \tool{} to find relevant bug reports for recommendation.
To that end, the idea is to first try to find existing Brave bug reports that are similar to the ones of FireFox to build a set of pairs of similar bug reports $<BR_{Firefox},BR_{Brave}>$.
Then, for each pair $<BR_{Firefox},BR_{Brave}>$, we rely on \tool to identify Brave app reviews that match $BR_{Firefox}$. 
However, we only consider Brave app reviews which precede the creation time of the corresponding $BR_{Brave}$.  
Finally, we manually check if the identified Brave app reviews match the corresponding Brave bug report $BR_{Brave}$. This would indicate that \tool would have been useful to automatically recommend the Firefox bug report as relevant to Brave. 

In practice, to build our set of similar bug reports, we randomly picked 3,000 bug reports from Firefox. 
By using a cosine similarity threshold of 0.91\footnote{We decided on 0.91 based on an empirical validation: the higher the similarity threshold is, the less pairs will be matched, and at the same time matched reviews are closer to bug reports. We decide on a  high threshold to maximize high quality results.}, we were able to identify 81 bug report pairs $<BR_{Firefox},BR_{Brave}>$ where the Firefox and Brave bug reports are highly similar. Note that this ground truth construction may be too conservative: there are possibly other Brave bug reports that are also semantically similar to a given Firefox bug report. 

Recall that the objective of \tool{} is to match bug reports to app reviews. In this case, we assess whether, for each pair of the ground truth, we can match the Firefox bug report with app reviews that are relevant (based on human expertise\footnote{This is the core challenge of \tool: to find semantic similarity between a bug report and an app review}) to the associated Brave report. If so, we can conclude that \tool{} would have been able to recommend the Brave bug report. 

For our experiments, we applied \tool on each of the 81 Firefox bug reports to match Brave app reviews. We retain only the top 3  matched reviews per bug report.
Overall, among the 81 top$_1$ reviews recommended by \tool{}, 21 could be confirmed as indeed semantically similar to the content in the Brave bug report associated in the ground truth pair. 

We further confirm 32 and 38 (at Top 2 and Top 3 respectively) app review matches.  
Table \ref{tb: ability} details our results by also providing the Accuracy and MRR scores. Overall, with Top 3 recommendations on a conservative ground truth, we reach almost 50\% hit ratio. 
Note that, while \tool{} matches Firefox bug reports with Brave app reviews, our effectiveness evaluation is to check whether the matched app reviews are semantically relevant for Brave bug reports. This is the practical and ultimate concern of our bug  recommendation scheme.

\begin{table}[]
\centering
\caption{Results of Acc@N and MRR@N.  {\normalfont "App review Hits" represents the number of times \tool{} matches the relevant app reviews associated to the ground truth bug report of Brave: this is a proxy for estimating that \tool{} would have been able to recommend the Firefox bug report as relevant to Brave.}}
\resizebox{0.7\linewidth}{!}{
\begin{tabular}{c|c|c|c}
\toprule
\begin{tabular}[c]{@{}c@{}}81 relevant bug pairs out of \\ 3K bugs from Firefox\end{tabular} & @1 & @2 & @3 \\ \midrule
App Review Hits                              & 21    & 32    & 38    \\ \hline
Value of Acc@N(\%)                       & 25.93 & 39.51 & 46.91 \\ \hline
Value of MRR@N(\%) & 25.93 & 26.50  & 35.19  \\ \bottomrule
\end{tabular}}
\label{tb: ability}
\end{table}

Table \ref{tb: rq2a} presents an extract of our set of 81 similar bug reports pairs by focusing on three pairs, the corresponding app user reviews and the result of our manual analysis. The full results are detailed on our Github repository\footnote{\url{https://github.com/BugRMSys/BugRMSys/blob/main/RQ1UserCase_firefox2brave.csv}}.

\find{{\bf \ding{45} Answer to RQ-1: }$\blacktriangleright$ The Acc@N and MRR@N values show that \tool{} is reasonably effective in matching relevant app reviews of a target app to bug reports from same-category apps in order to drive bug recommendation. $\blacktriangleleft$ }


\subsection{[RQ-2]: Feasibility of \tool{}}
\subsubsection{Bug recommendation in the wild}
We conduct extensive execution of \tool{} on data from 20 apps in 9 categories to recommend bugs. 
Due to space limitation, we report in Table~\ref{tb: top10apps} the statistics of bugs recommended by \tool{} for the top$_{10}$ apps having the most recommended bugs.
We note that, thanks to \tool{} app-review driven approach, the collaborative bug finding allows to sift between a few hundreds to a few thousands bug reports from same-category apps in order to recommend\footnote{We set a high similarity threshold at 0.9. This value can be fine-tuned following the practitioner's objectives.} only a few (1.63\%) of bug reports as being relevant to the target apps. 

\begin{table}[hhh]
\caption{Ranked apps based on the number of potential bugs.}
\centering
\resizebox{1\linewidth}{!}{
\begin{tabular}{l|r|r|r}
\toprule
Target  & \# of bug reports searched &\# app reviews & \# recommended \\ 
app & (from same category apps) & (for the target app) &   bugs \\ \hline
Brave  & 10,000&    10,000            &    208               \\ \hline
Nextcloud  & 3,489 &    1,143            &    147               \\ \hline
Wire & 10,000&    2,515            &    75               \\ \hline
VLC  & 599&    5,556            &    59               \\ \hline
Firefox  & 10,000&    10,000            &    52               \\ \hline
Dolphin    &    2,437        & 2,498   &    44               \\ \hline
Wordpress    &   4,139         & 2,714   &    44               \\ \hline
PPSSPP    &  1,666          & 6,846   &  39                 \\ \hline
Mupen64Plus    &  2,437          & 1,401   &  39                 \\ \hline
Fbreader    &    726        & 3,914   & 33                  \\ \bottomrule
\end{tabular}}
\label{tb: top10apps}
\end{table}


Since each recommended bug is found by correlating information in its app reviews, we propose to estimate the potential time gain \tool{} has brought by highlighting the buggy behaviour users complained about in unofficial channels. We compute the distribution of time elapsed since the app review creation date and the \tool{} bug recommendation date.
On average, specially for Firefox and Brave in Table~\ref{tb: top10apps}, the app reviews were created 22.2 and 33 days before we submit the bug reports, respectively.



\find{{\bf \ding{45} Answer to RQ-2.a }$\blacktriangleright$ \tool{} can help triage bug reports from same-category apps to recommend a reasonable number of bugs in a target app. We also show that \tool{} helps to highlight bugs that could have remained overlooked for a long time in app reviews, {e.g.} 22.2 days earlier for Firefox, and 33 days earlier for Brave.
$\blacktriangleleft$}

To assess whether the bugs recommended by \tool are real bugs, 
we further reproduce each recommended bugs manually, and submit the successfully reproduced bugs in the issue tracker of the related app. 
Since the manual reproducing work requires extensive efforts, we will focus our investigation on four popular apps: Wire, Brave, FireFox and Nextcloud.

The numbers of recommended bugs, (successfully) reproduced bugs (these reproduced bugs are selected with their similarities, and the top-44 most similar ones are selected), and confirmed or fixed bugs are presented in Table~\ref{tb: newBugs}. The details of the reported bugs can be found on our repository\footnote{\url{https://github.com/BugRMSys/BugRMSys/blob/main/new_bugs.md}}.
We remind that we can recommend a bug report from app {\tt A} (the first column in Table~\ref{tb: newBugs}) as relevant to the target app {\tt B} when \tool matches the bug report of {\tt A} with user app reviews from the target app {\tt B}. 
We then use the ``steps to reproduce'' present in the bug report of {\tt A}, as well as the information present in the app reviews of app {\tt B} to reproduce the bug in the target app {\tt B}. 
Finally, for each bug that has been successfully reproduced in the target app {\tt B}, we submit the bug in the issue tracker of the app. As shown in the the last column of Table~\ref{tb: newBugs}, six bugs have been already confirmed or fixed by the developers before this submission. 


\begin{table*}[hhh]
\caption{Previously unknown bugs detected with \tool.}
\centering
\resizebox{1\linewidth}{!}
{
\begin{threeparttable}
\begin{tabular}{l|l|r|r|r|r|r|r}
\toprule
Input &Target  & \# of bug reports  &\# app reviews & \# recommended & \# of reproduction & \# successully  & \# replied, confirmed or \\ 
app& app & (input apps) & ( target app) &   bugs &  attempts & reproduced bugs & fixed bugs\\ \midrule

Signal & Wire & 10,000&    2,515          & 75 & 12 &   2  &  2, [\ding{52} 2]       \\ \hline
Firefox & Brave    &  10,000       & 10,000 & 208 &44 &  9    &     (1),2, [\ding{52} 2]      \\ \hline
Brave & Firefox    &  10,000          & 10,000   &52 &24 &   4     &    (1), 1      \\ \hline
Owncloud & Nextcloud &     3,489      &    1,143 &  147 &12  & 5 &    (1), 1         \\ \midrule
 &  &            &  &  482 &  90   & 20  & (3), 6, [\ding{52} 4]           \\ \bottomrule
\end{tabular}
{$^\ast$``(\#)'' means the number of reported bugs replied by developers but not confirmed or fixed by them.\\ ``[\ding{52} \#]'' means the number of reported bugs confirmed and fixed by developers.}
\end{threeparttable}
}
\label{tb: newBugs}
\end{table*}

\begin{table*}[]
\caption{Extract of our Ground Truth Dataset, Corresponding relevant app reviews, and Manual check result.}
\resizebox{1\textwidth}{!}{
\begin{tabular}{l|l|l|c}
\toprule
\multicolumn{2}{c|}{Existing Bug Reports (input, target)} &
  {TOP 3 RELEVANT REVIEWS in Brave \emph{(creation time always prior to the corresponding Brave bug report) }} &
  MANUAL \\ \hline
{\multirow{3}{*}{\makecell[l]{App: Firefox\\ Data: 2020-08-21\\ Reports: Download does not  \\ work  on a custom  tab (Slack)}}} &
  \multirow{3}{*}{\makecell[l]{App: Brave\\ Date: 2021-03-20\\ Report: Download {[}Status\\ Bar{]} Improvement}} &
  \makecell[l]{Date: 2020-08-22\\ Review: Downloader is very bad....pls increase and more work on download manager} &
  TRUE \\ \cline{3-4} 
 &
   &
  \makecell[l]{Date: 2020-08-10\\ Review: There is no download option in this could u pls update on this issue} &
  TRUE \\ \cline{3-4} 
 &
   &
  \makecell[l]{Date: 2020-10-18\\ Review:...it does not allow to manually add download tasks} &
  TRUE \\ \hline
{\multirow{3}{*}{\makecell[l]{App: Firefox\\ Data: 2020-06-17\\ Reports: Report clickbait sites, \\ Protect user privacy}}} &
  \multirow{3}{*}{\makecell[l]{App: Brave\\ Date: 2020-06-22\\ Report: Warn users about \\ insecure Facebook and \\ Google privacy settings}} &
  \makecell[l]{Date: 2020-06-08\\ Review: Only browser that cheats about privacy.  All claims about user privacy are bogus...} &
  TRUE \\ \cline{3-4} 
 &
   &
  \makecell[l]{Date: 2020-11-01\\ Review: Extremely convoluted privacy practices. They advocate for privacy but allow certain  creepy sites ...} &
  FALSE \\ \cline{3-4} 
 &
   &
  \makecell[l]{Date: 2020-12-31\\ Review: Good privacy app. It doesn't prevent websites from annoying redirections} &
  FALSE \\ \hline
{\multirow{3}{*}{\makecell[l]{App: Firefox\\ Data: 2020-08-28\\ Reports: Invalid URLs can \\ be bookmarked and they\\  crash the browser}}} &
  \multirow{3}{*}{\makecell[l]{App: Brave\\ Date: 2020-11-06\\ Reports: Clicking URLs \\ outside of Brave opens a\\ blank browser window\\ with no URL}} &
  \makecell[l]{Date: 2020-06-11\\  Review: "..., has come under fire for automatically  redirecting URLs typed into the browser's  address bar ...} &
  FALSE \\ \cline{3-4} 
 &
   &
  \makecell[l]{Date: 2020-10-20\\ Review: One of the best browsers Imo. Only wish I could  set links on the brave homepage manually...} &
  FALSE \\ \cline{3-4} 
 &
   &
  \makecell[l]{Date: 2020-10-13\\ Review: Its a good browser sometimes it reload all tabs when i open newly} &
  FALSE \\ \bottomrule
\end{tabular}}
\label{tb: rq2a}
\end{table*}

\find{{\bf \ding{45} Answer to RQ-2.b: }$\blacktriangleright$  Thanks to \tool{}, we found 20  new bugs efficiently (i.e., within a short period of time), among which six have been confirmed or fixed by developers. These results demonstrate that \tool{} is relevant for exploring news bugs using its app reviews driven collaborative bug finding scheme. Since there is no training involved in \tool{},  it can be readily applied to various apps from various categories to help developers find bugs that have not yet been officially reported but which users may have witnessed already. $\blacktriangleleft$ }

\subsection{[RQ-3]: \tool{} vs Prior Works}
Ideally, we should evaluate the performance of \tool{} in comparison with prior works dealing with bug recommendations based on bug reports. There are two state of the art approaches, DeepMatcher~\cite{haering2021automatically} and Bugine~\cite{tan2020collaborative}, which are closely related.

DeepMatcher and \tool{} both match app reviews with bug reports based on text embedding using pre-trained DistilBERT. Experimentally, we compare DeepMatcher against the \tool{} by considering the ground truth data built for RQ-1 (cf. examples in Table~\ref{tb: rq2a}): we propose to manually check whether the matched reviews with both approaches are relevant or not. While all reviews matched by \tool{} are relevant to buggy behaviour, we observe that DeepMatcher only achieves a F1-score of 71\% in filtering useful reviews. We postulate that \tool{} performs better partly because it implements a focused collaborative bug finding approach where the matching is done on bug reports of same-category apps.


We also compare against Bugine, which also performs collaborative bug finding. We differ however as Bugine limits the matching to cases where apps have the UI/components (while we consider apps from teh same categories). We further consider app reviews to drive bug recommendation. 

In the remainder, we further elaborate on the specific differences that prevent comparison between prior works and \tool{}. These differences relate to three aspects: 
(1) Differences in input, output, and workflow; (2) Differences in automation level; (3) Performance in new bug finding.

\subsubsection{Differences in input, output, and workflow}

DeepMatcher employs user reviews for App B as input and recommends relevant bug reports for App B. The workflow is: App review $\rightarrow$ Problem report $\rightarrow$ Matched relevant bug reports. They only evaluate their tool on existing bugs instead of exploring new bugs. Furthermore, DeepMatcher does not leverage experience from other apps when investigating a target app. Their approach further suffers from the redundancy problem in app reviews.

Bugine employs issues in apps with same UI components as their database. They focus on building a automatic test generation from bug reproduction steps and run the test on target app with manual check. Bugine has been used to explore new bugs successfully. However, there is a great limitation in this approach: it only considers app issues with same-UI components into consideration, which can reduce the feasibility of learning from other apps.


\subsubsection{Difference in automation level}
Different from DeepMatcher, \tool{} will not process a large number of bug reports: we focus on same-category apps to match relevant reviews of App B. After manual check, we have verified that when we feed a bug report into \tool{}, the matched reviews are 100\% related to  some bugs. By building on same category apps (i.e., with similar functionality and usage steps) reproduction and localization of bugs is eased.

For Bugine, finding apps with same-UI components is a time-consuming task. In addition, using issues from same-UI apps makes it hard to transfer the learned expertience to other types of issues.

\subsubsection{Performance in bug finding}

The ability of DeepMatcher to find new bugs has not been evaluated. Bugine reported having found 34 new bugs in  5 evaluated apps.
With \tool{}, within a week, we were able to recommend, reproduce and identify 20 new bugs across 6 apps. 4 such bugs are already fixed by the app developers.

\find{{\bf \ding{45} Answer to RQ-3:}$\blacktriangleright$ \tool{}, in comparison to Deepmatcher, is effective for filtering out relevant app reviews. Compared to Bugine, \tool{} is scalable and can be applied to a larger range of bug types, while avoiding duplication of recommending bugs that were already reported in the target app. $\blacktriangleleft$ }

%% file: 6.Discussion.tex
\section{Discussion}\label{sec:discussion}
\subsection{Failures to reproduce recommended bugs}
As illustrated previously, some of the reproduction attempts on the bug reports recommended by \tool{}. lead to failures There are various reasons that explain such failures without suggesting that the recommended bug is not relevant. Prior studies have already largely elaborated on this difficulty to reproduce bugs: In Han's work~\cite{han2019reproducing}, an extensive classification of 8 categories of root causes for failed reproductions is provided: hardware dependency, operating system dependency, component dependency, unavailable source code, compilation error, installation error, missing step, and lack of symptom. Our failures causes span across these categories. 

\subsection{Threats to validity}


Our design, implementation and evaluation of \tool{} carries some threats to validity. First, when we are building the ground truth, we manually check whether the reviews are meaningful. Therefore, the ground truth may be biased by our own experience. Second, \tool{} is not fully automated, i.e., manual effort is still needed when we reproduce from the recommended bug reports. Consequently, the success rate of reproduction could be dependent on the developing experience of individual developer. 

%% file: 2.Relatedwork.tex
\section{Related Work}\label{sec:related}
\textbf{Collaborative Experience Sharing:}
Collaborative programming is common in the development of open source software. Consequently, similar bugs can emerge across different projects.
Other researches attempted to leverage this fact to Recommend, Reproduce, and Repair inter-project bugs. For instance, the detection of duplicate bug reports has been studied in localizing fault of software~\cite{sun2011towards,wang2008approach}. Specifically, Yang~{\em et al.}~\cite{yang2016combining} combined the information retrieval technique and word embedding technique to process the detailed information of bug reports to recommend similar bugs. On the other hand, Tan {\em et al.}~\cite{tan2020collaborative} use three collaborative sources for bug finding: (1) bugs from the same programmer across different projects, (2) bugs from manually searching for bug reports in GitHub repositories, (3) bugs from a bug recommendation system. Based on these shared experiences, they explored the concept of collaborative bug finding on improving the teaching of software testing courses. In the experience-based collaborative learning of crowd-sourcing, Mao~{\em et al.}~\cite{mao2017crowd}'s experimental results show that the generated replicable test scripts from crowd-based testing can improve the coverage attainment for automated mobile testing.

\textbf{Recommendation of Bug Reports Based on App Reviews:}
The importance of app reviews in App vendors has been comprehensively demonstrated ~\cite{oh2013facilitating,Ashley2015Survey}. Leaving the app reviews not to be addressed is harmful to the experience of the users and rating of the app, and further lead to uninstallation of the app \cite{hassan2018studying}. To maintain the evolution of the app, researchers started to leverage app reviews. For instance, Gao {\em et al.}~\cite{gao2019automating} developed a novel approach to automatically generate proper responses to the app reviews in Google Play. However, this approach mainly try to (1) soothe bad emotion of users, and (2) collect detailed user experience, but not to discover potential bugs in advance. Tan {\em et al.}~\cite{tan2020collaborative} designed an approach to find bugs for Android apps. Their pipeline is to retrieve bugs in other similar apps that may also exist in the current app. This work validates the feasibility of searching for bugs in other projects to identify new bugs mentioned by app reviews. Afterwards, Marlo {\em et al.}~\cite{haering2021automatically} try to match bug reports with related app reviews to discover bugs by filling the gap of different languages between app reviews written by non-technical users and bug reports proposed by professional developers.


%% file: 7.Conclusion.tex
\section{Conclusion and Future Work}\label{sec:conclusion}

In this paper, we introduce \tool{}, a tool-supported approach for app reviews driven collaborative bug finding. Given a target app {\tt B}, \tool{} builds on the development experience of app {\tt A} to identify bug reports in {\tt A} that match app reviews of {\tt B}. If such bug reports exist, they are considered as candidate for recommending bugs to the target app {\tt B}.
To that end, \tool{} implements an embedding procedure to represent bug reports and app reviews text, and use cosine similarity to decide on matching similarity scores. Once bugs are recommend, we experimentally attempt reproduction to confirm the detection of new bugs in the target app.
Our experimental results on free and open source apps in various categories show that \tool{} is effective, scales to a variety of bug types, and does not yield too many irrelevant app review matches.  
Overall, with \tool{}, we already successfully reproduced 20 new bugs in 6 apps across 3 categories. Several of these bugs have been acknowledge by the apps development communities and some have even already been fixed.

In future, we plan to address the question of automating the reproduction phase in order to scale the collaborative bug finding approach towards further increasing its practicality in  real-world debugging scenarios.

\section{Conflict of Interest Statement}
The authors declare that they have no conflict of interest.
\section{Data availability}
The datasets and code used in this work are available in the following link:
\begin{center}
    {\bf\url{https://zenodo.org/record/7520604#.Y71lnezML0o}}
\end{center}